\documentclass[11pt]{article}
\usepackage{graphicx,subfigure}
\input{amssym.def}

\begin{document}

\title{Self-similarity of Farey Staircases}
\author {Grynberg, S.\footnote{e-mail: sebgryn@fi.uba.ar} and Piacquadio, M. \\Departamento de Matem\'atica\\Facultad de Ingenier\'{\i}a\\ Universidad de Buenos Aires\\Paseo Col\'on 850, (1063)\\Buenos Aires, Argentina.}
\date{}
\maketitle

\begin{abstract}
We study Cantor Staircases in physics that have the Farey-Brocot arrangement for the $\frac QP$ rational heights of stability intervals $I(\frac QP)$, and such that the length of $I(\frac QP)$ is a convex function of $\frac 1P$. Circle map staircases and the magnetization function fall in this category. We show that the fractal sets $\Omega$ underlying these staircases are connected with key sets in Number Theory via their $(\alpha, f(\alpha))$ multifractal decomposition spectra. It follows that such sets $\Omega$ are self similar when the usual (Euclidean) measure is replaced by the hyperbolic measure induced by the Farey-Brocot partition.
\end{abstract}

\section{Introduction}

A Cantor staircase is an increasing continuous function from $[a,b]$ to $[0,1]$, $y=g(x)$, with zero derivative almost everywhere, constant on the so-called intervals of resonance or stability $\Delta x_k, k\in\Bbb N$. The complement in $[a,b]$ of $\bigcup_{k\in\Bbb N}\Delta x_k$ is a totally disconnected Cantordust set $\Omega$.

Cantor staircases are frequently observed in empirical physics, and their universal properties are of great interest. These staircases are naturally associated with a Cantordust set $\Omega$. Such an $\Omega$ reflects the particular physical problem under study.

Such Cantordusts can be studied with the tools provided by Number Theory and the multidimensional $(\alpha, f(\alpha))$ decomposition of a fractal set $\Omega$.

\subsection{The Ising Model and the Circle Map}
\begin{enumerate}
\item[(a)] 
{\bf The Ising model.}
Bruinsma and Bak [1983] studied the one dimensional Ising model with convex long-range antiferromagnetic interaction. Only "up" spins interact, their interaction being given by a convex function depending on a parameter $a>1$, $a$ is the strength of the interaction. Let $H$ be the applied magnetic field and $q$ the proportion of up-spins. At the critical temperature $T=0$ the phase diagram $q=g(-H)$ exhibits a Cantor staircase.

With $\Delta H$ we will denote the intervals of resonance or stability of the staircase $q=g(-H)$. {\it Par abus de langage} $\Delta H$ will be the corresponding stairstep as well as its length. $\Delta H(\frac QP)$ means: the stairstep of rational height $\frac QP$ in the staircase. Bruinsma and Bak state that $\Delta H(\frac QP)$ depends only on $P$, and that
\begin{equation}
\label{fases}
\gamma\left(\Delta H\left(\frac QP\right)\right)^{\frac 1{a+1}}\cong\frac 1P
\end{equation}
where $\gamma$ is a constant depending only on $a$, the interaction strength.
\item[(b)]
{\bf The Circle Map.}
The simple sine circle map 
\begin{equation}
\theta_{n+1}=\theta_n+\omega+\frac{\sin(2\pi\theta_n)}{2\pi}
\end{equation}
is one of the simplest models describing systems with two competing frequencies --e.g. the forced pendulum. Here $\theta$ is the angle formed by the vertical and the pendulum; $n$ is the discretized time  variable; $\omega$ represents the frequency of the system in the absence of the nonlinear term given by the sine function. Let $W$ be the winding number corresponding to the average
$$\lim\limits_{n\to\infty}\frac{\theta_n}{n}.$$
The graph of the function $W=g(\omega)$ is a well known Cantor staircase. With $\Delta\omega$ we denote its intervals of resonance, as well as the corresponding stairsteps and their length.
\end{enumerate}

\subsection{Universal Properties of these two Cantor Staircases}
\begin{enumerate}
\item[(a)] 
{\bf Farey-Brocot.}
Let $y=g(x)$ be any of these two Cantor staircases. Let $\Delta x$ and $\Delta x'$ be two intervals of resonance. Let us further suppose that each $\Delta x''$ in the gap between $\Delta x$ and $\Delta x'$ has size smaller than those of both $\Delta x$ and $\Delta x'$. Let $\frac QP$ be $g(x)$ when $x\in\Delta x$, and let $\frac{Q'}{P'}$ be $g(x')$ when $x'\in\Delta x'$. Then, if $\Delta x''$ is the largest interval in the gap, and if $x''\in\Delta x''$, one has $g(x'')=\frac{Q''}{P''}=\frac{Q+Q'}{P+P'}$. 
\item[(b)]
{\bf Hausdorff dimension.}
Let us recall that the Cantordust $\Omega$ naturally associated with a Cantor staircase is the complement --in the domain $[a,b]$ of $y=g(x)$-- of the union of the intervals of resonance. For each such $\Omega$ associated with the staircases quoted above, we have $d_H(\Omega)\in (0,1)$, where $d_H$ is the Hausdorff dimension... i.e. $\Omega$ is, strictly speaking, a fractal set. For the Ising Model case, Bruinsma and Bak estimated $d_H(\Omega)$, and for the Circle Map the result $d_H(\Omega)=0.87...$ is a known universal number.
\end{enumerate}

\subsection{The Tools Provided by Number Theory and the Multifractal Spectrum $(\alpha, f(\alpha))$}
\begin{enumerate}
\item[(a)]
{\bf Number Theory.} The problem of approximating irrational numbers by rational ones is a key subject in Number Theory. Let $i\in(0,1)$ be an irrational number, $Q$ and $P$ in $\Bbb N$. Both in Number Theory and its applications, the approximation of $i$ by different rationals $\frac QP$ is given by the study of the distance $\left|i-\frac QP\right|$. Farey-Brocot sequences and continued fractions provide the tools to study the evolution (behaviour, dynamics) of this distance.
\begin{enumerate}
\item [(1)]
{\bf Farey-Brocot $(F-B)$ sequences.}
Farey-Brocot sequences $(F-B)_n$ with $n\in\Bbb N$ are defined thus:
$(F-B)_1=\left\{\frac01,\frac11\right\}$, $(F-B)_2=\left\{\frac01,\frac{0+1}{1+1},\frac11\right\}=\left\{\frac01,\frac12,\frac11\right\}$, and, once $(F-B)_n$ is defined, we will define $(F-B)_{n+1}$ by interpolating as follows: we take each consecutive pair $\frac QP$ and $\frac{Q'}{P'}$ in $(F-B)_n$, $\frac QP<\frac{Q'}{P'}$, and we interpolate $\frac{Q+Q'}{P+P'}$ between $\frac QP$ and $\frac{Q'}{P'}$, adding the fraction $\frac{Q+Q'}{P+P'}$ to $(F-B)_n$ in order to make $(F-B)_{n+1}$. The first $(F-B)'s$ are:
$$(F-B)_2=\left\{\frac01,\frac12,\frac11\right\}$$
$$(F-B)_3=\left\{\frac01,\frac13,\frac12,\frac23,\frac11\right\}$$
$$(F-B)_4=\left\{\frac01,\frac14,\frac13,\frac25,\frac12,\frac35,\frac23,\frac34,\frac11\right\}$$
...and so on.
\item[(2)]
{\bf Continued Fractions.}
Any irrational number $i\in (0,1)$ can be written uniquely as
$$
i={1\over\displaystyle a_1+{1\over\displaystyle a_2+{1\over\displaystyle  a_3+{_{~~\displaystyle\ddots}}}}}
$$
$$=[a_1,a_2, a_3,...],\;a_n\in\Bbb N.$$
This infinite continued fraction $i$ when cut off at $n$, i.e. $$[a_1,a_2,..., a_n],$$
is a rational number $\frac{Q_n}{P_n}$ that well approximates $i$ as $n\to\infty$, which means that
$$\left|i-\frac{Q_n}{P_n}\right|<\frac{1}{P_n^2},$$
$n\in\Bbb N$.
\item[(3)]
{\bf The relationship between continued fractions and $(F-B)$ sequences.}
Let $i=[a_1,a_2,..., a_n,...]$. $\frac{Q_1}{P_1}$ is in the $a_1$-th $(F-B)$ sequence, and in general, $\frac{Q_n}{P_n}$ is found in an $(F-B)$ sequence $a_n$ steps ahead of the $(F-B)$ sequence in which $\frac{Q_{n-1}}{P_{n-1}}$ appears.
\end{enumerate}
\item[(b)]
{\bf The multifractal or multidimensional $(\alpha, f(\alpha))$ of a fractal set $\Omega$.}
Multifractal decomposition is a useful tool [Halsey et al., 1986], first conceived by physicists, to study a fractal $\Omega$ with a somewhat irregular geometric configuration, such as the Cantordust underlying our $y=g(x)$ staircases. Let us consider the ternary set $K\subset [0,1]$ of Cantor. $K$ is an example of a geometrically very regular fractal set. K is obtained applying contractive transformations $T_1$ and $T_2$ to $[0,1]=I$; $T_1(I)=[0,\frac13]$ and $T_2(I)=[\frac23,1]$. Successive iterations of $T_1$ and $T_2$ produce $K$. Words of $k$ letters $T_i\; (i=1,2)$, yield $2^k$ segments in $I$, which constitute the $k^{th}$ approximation to $K$. The contractors associated with $T_1$ and $T_2$ are both $\frac13$; i.e. $K$ is a $\frac13-\frac13$ fractal. We provide $K$ with a probability measure $p$ such that $p(T_1(I)\cap K)=p_1=\frac12$ and $p(T_2(I)\cap K)=p_2=\frac12$. All $2^k$ segments $I^k$ in the $k^{th}$ partition fulfill $p(I^k\cap K)=\frac{1}{2^k}$. Let $\left|I^k\right|$ be the length of segments $I^k$. The equation for the $\alpha$-index of concentration of $I^k$:  $p(I^k\cap K)=\left|I^k\right|^{\alpha}$ will yield $\alpha(I^k)=\frac{\log 2}{\log 3}$, a number independent of $k$. Therefore any point $x\in K$, via its sequence of nested intervals $I^k=I^k(x), k\in\Bbb N$, will inherit a concentration $\alpha(x)=\frac{\log 2}{\log 3}$. Let us consider a more irregular Cantordust $\Omega$, say, $\Omega$ is a $\frac13-\frac14$ fractal with $p_1=p_2=\frac12$. Intervals $I^k$ in the $k$-partition do not necessarily share the same length, though we still have $p(I^k\cap \Omega)=\frac{1}{2^k}$. The concentration $\alpha$, therefore, varies from segment to segment... hence, from point in $\Omega$ to point in $\Omega$. Let $\Omega_{\alpha}$ be the set of all $x$ in $\Omega$ sharing the same concentration $\alpha$. The multifractal or multidimensional spectrum $f(\alpha)$ is, by definition, $d_H(\Omega_{\alpha})$. The Cantordusts $\Omega$ underlying the Cantor staircases quoted above are much more irregular than any $\frac1m-\frac1n$ fractal.
\end{enumerate}
\subsection{On Continued Fractions and Jarn\'{\i}k Classes}
\begin{enumerate}
\item[(a)]  
Throughout this paper we will need a number of properties about continued fractions.

Let $i=[a_1,a_2, ...,a_n,...]$, 
$a_n\in\Bbb N$, be an irrational number. 
Let us recall that 
$[a_1, a_2, ..., a_n]=$
$\frac{Q_n}{P_n}$ well approximates $i$ as $n\to \infty$; 
$\left|i-\frac{Q_n}{P_n}\right|<\frac{1}{P_n^2}$, 
$n\in\Bbb N$.

The denominator $P_n$ tends to $\infty$ because we have $P_{n+1}=a_{n+1}P_n+P_{n-1}$; $P_{-1}=0$; $P_0=1$.

The polynomial $P_n=P_n(a_1, a_2,...,a_n)$ has $F_n$ monomials, where $F_n$ is the $n^{th}$ Fibonacci number, $F_n=\frac{1}{\sqrt5}(\phi^n-\phi^{-n})\simeq c\phi^n$, $\phi$ the golden mean. Therefore, we have that $P_n\geq F_n$.
 
\item[(b)]
What we should recall about Jarn\'{\i}k classes.

The set of irrationals $i$ for which $\left|i-\frac{Q}{P}\right|<\frac{1}{P^{\beta}}$
(for infinite values of $P\in\Bbb N$) is called the $J_{\beta}$ class of Jarn\'{\i}k, here $\beta\geq 2$ [Falconer, 1990]. 

Clearly $J_{\beta_1}\supset J_{\beta_2}$ when $\beta_1<\beta_2$.

Results by Dirichlet and Jarn\'{\i}k [Falconer, 1990] show that $d_H(J_{\beta})=\frac{2}{\beta}$, $\beta\geq 2$.  
\end{enumerate}

\section{Cantor Staircases and the Ising Model}
Let $y=g(x)$ be the staircase describing the magnetization process, $x=-H$ and $y=q$ the proportion of "up" spins as described in section 1.

In a previous paper [Piacquadio and Grynberg, 1998] we studied the size of intervals of stability $\Delta H$. We studied the size of steps $\Delta H$ near points of irrational height $i$ in the staircase. What follows is a brief sketch of the contents of that paper.

Each step $\Delta H$ has rational height $\frac QP$; we studied the size of steps $\Delta H(\frac QP)$ when $\frac QP$ well approximates a certain irrational $i$. Notice that Eq.(\ref{fases}) already gives the value of the length of $\Delta H(\frac QP)$. If $\frac{Q_n}{P_n}=[a_1, a_2,...,a_n]$, $i=[a_1, a_2,...,a_n,...]$, we will deal with $\Delta H(\frac{Q_n}{P_n})$.

In what follows, and for short, with the same symbol $\Delta H$ we will refer to the interval of resonance in the $-H$ axis, to the corresponding stairstep in the graph of $y=g(-H)$, and to the length of either. We trust that the context will avoid confusion.

Given a certain irrational $i$ there exists a unique point $A_i$ of height $i$ in the Cantor staircase and, given a certain small $\varepsilon>0$, there is an infinity of stairsteps $\Delta H$ at no-bigger-than-$\varepsilon$ distance of the point $A_i$. Let $\Delta H_{i,\varepsilon}$ be the largest of them.

Clearly, as $\varepsilon$ goes to zero, so does $\Delta H_{i,\varepsilon}$. For a fixed value of $i$, we are interested in computing $\Delta H_{i,\varepsilon}$ as a diminishing function of $\varepsilon$. Now, such a stairstep $\Delta H_{i,\varepsilon}$ has rational height in the staircase. We show that
\begin{enumerate}
\item[(a)]
for the sake of our computations, $\varepsilon=\varepsilon_n=\frac{1}{P_nP_{n+1}}$ is an appropriate choice of an $\varepsilon$ going to zero.
\item[(b)]
for such an $\varepsilon_n=\varepsilon$ the stairstep $\Delta H_{i,\varepsilon}$ has rational height $\frac{Q_n}{P_n}$, that is, we have $\Delta H_{i,\varepsilon}=\Delta H(\frac{Q_n}{P_n})$ in the notation of Bruinsma and Bak.
\end{enumerate}

In order to study the behaviour of $\Delta H_{i,\varepsilon}$ as a function of $\varepsilon$, we introduced in [Piacquadio and Grynberg, 1998] the notion of "type": Let $k\in\Bbb N$, $k\geq 2$.

If $i$ is such that $\frac{1}{P_n}$ fulfills
\begin{equation}
\label{tipounosobrek}
\lim\limits_{\varepsilon_n\to 0}\frac{\frac{1}{P_n}}{\varepsilon_n^{\frac 1k +\delta}}=\infty\;\; \mbox{and}\;\; \lim\limits_{\varepsilon_n\to 0}\frac{\frac{1}{P_n}}{\varepsilon_n^{\frac 1k -\delta}}=0
\end{equation}
for an arbitrary small $\delta>0$, we say that $\frac{1}{P_n}$ goes to zero strictly like $\varepsilon_n^{\frac 1k}$, we call $i$ "type $\frac 1k$", we place every such $i$ in a pigeonhole $G_k$, $k\in\Bbb N$, $k\geq 2$, and we write $\frac{1}{P_n}$ as $\kappa_n\varepsilon_n^{\frac 1k}$, $\kappa_n$ bounded or not.

We show that, if $i\in G_k$, we have $\Delta H_{i,\varepsilon}$ determined by $\sqrt{\varepsilon}^{d_H(J_k)}$; that is, we have $\frac{1}{P_n}$ going to zero strictly as a power, the base of which is $\sqrt{\varepsilon}=\sqrt{\varepsilon_n}$ (i.e. the smallest possible diminishing function), and the exponent is $d_H(J_k)$.

Essentially, the size of steps $\Delta H_{i,\varepsilon}$ depends on the Jarn\'{\i}k classes to which $i$ belongs.

The shortest intervals $\Delta H_{i,\varepsilon_n}$ correspond to $i=[1,...,1,...]=\phi^{-1}=\phi-1$, where $\phi= \frac{1+\sqrt5}{2}$, the golden mean. In that case $\Delta H_{i,\varepsilon_n}=\Delta H_{\phi,\varepsilon_n}$ is given by $\sqrt{\varepsilon_n}$ multiplied by a coefficient $\kappa\cong 1.27$. Already for $i=[2,...,2,...]=s$, the silver mean, we have $\Delta H_{s,\varepsilon_n}$ given by $\sqrt{\varepsilon_n}$ multiplied by a coefficient strictly larger than $1.27$. Irrationals $\phi$ and $s$ are, of course, in $G_2$. So are $[1,2,...,n,...]$ and $[1^2,2^2,...,n^2,...]$. If the growth of $a_n$ is sufficiently accelerated, we have that $i=[a_1,a_2,...,a_n,...]$ may be in $G_3$, or in $G_4$,...

Since we have $d_H(G_k)\leq\frac 2k=d_H(J_k)$, $k\geq 2$, $k\in\Bbb N$, the dimension of $G_k$ diminishes as $k$ grows.

We had "disjointed" the $J_k$ classes ($J_2\supset J_3\supset\dots\supset J_k\supset J_{k+1}\supset\dots$) in disjoint rings $R_k=J_k-J_{k+1}$ and we had promised to show that $G_k\subset R_k$, $\forall k\in\Bbb N$, $k\geq 2$.

When we exhaust all rings $R_k$ we are left with irrationals $i\in J_{\infty}$ --i.e. the class of irrationals with ultrarapid growth of $a_n$. Such class has zero Hausdorff dimension, and is contained in every $J_k$, $k\geq 2$. The condition $\lim\limits_{n\to\infty}\frac{\ln P_{n+1}}{\ln P_n}=k-1$ ensured $i\in G_k$. The condition $\lim\limits_{n\to\infty}\frac{\ln P_{n+1}}{\ln P_n}=\infty$ ensured $i\in J_{\infty}$. 

\section{The Shells $G_{\beta}$}
In fact, $R_k=J_k-J_{k+1}$, $k\in\Bbb N$, is a very "thick" ring, i.e. the case $k\in\Bbb N$ is very far from being a general one. In this section we will prove

{\bf Theorem.} Let $\beta\in\Bbb R$, $\beta\geq 2$. Let $G_{\beta}$ be defined by Eq. (\ref{tipounosobrek}) with $\beta\in\Bbb R$ in place of $k\in\Bbb N$. Then $$G_{\beta}=\bigcap\limits_{\beta-2\geq\theta>0}J_{\beta-\theta}- \bigcup\limits_{\delta>0}J_{\beta+\delta},$$
and we have $\Delta H_{i,\varepsilon}$ given by $\sqrt{\varepsilon}^{d_H(G_{\beta})}=\sqrt{\varepsilon}^{d_H(J_{\beta})}$.

We will use two known properties of continued fractions:

{\bf Property 1.} Let $i$ be an irrational number, $i\in (0,1)$, and let $\frac rs \in\Bbb Q$. If
\begin{equation}
\label{Hardy}
\left|i-\frac rs\right|<\frac{1}{2s^2},
\end{equation} 
then $\frac rs=\frac{Q_n}{P_n}$, for some $n\in\Bbb N$.

{\bf Property 2.} Let $i$ be an irrational number, and $\frac{Q_n}{P_n}$ be a $n$-approximant to $i$. We then have
\begin{equation}
\label{app1}
\frac{1}{P_n(P_n+P_{n+1})}<\left|i-\frac{Q_n}{P_n}\right|<\frac{1}{P_nP_{n+1}}.
\end{equation}
We need three Claims: 

{\bf Claim 1.} Let $i\in (0,1)$ be an irrational number, let $\theta\in (0,1)$, let $\frac rs \in\Bbb Q$, and $\beta\geq 2$. Then we have that --except for a finite number of rationals $\frac rs$-- 
$$\left|i-\frac rs\right|<\frac{1}{s^{\beta+\theta}}$$
implies: $\frac rs$ is a $\frac{Q_n}{P_n}$ for some $n\in\Bbb N$.

{\bf Claim 2.} If $\limsup\limits_n\frac{\ln P_{n+1}}{\ln P_n}=\beta-1$, $\beta\geq 2$, and $P_n$ as above, then for any $\delta>0$ there exists $n_{\delta}\in\Bbb N$ such that
\begin{equation}
\label{claim2}
P_n(P_n+P_{n+1})<P_n^{\beta+\delta}
\end{equation}
$\forall n\geq n_{\delta}$.

{\bf Claim 3.} If $\limsup\limits_n\frac{\ln P_{n+1}}{\ln P_n}=\beta-1$, $\beta\geq 2$, then for any $\theta\in (0,1)$ there exists a sequence of naturals numbers $n_j=n_j(\theta)$ such that
\begin{equation}
\label{claim3}
P_{n_j}^{\beta-\theta}<P_{n_j}P_{n_j+1}.
\end{equation}

Let us prove the Theorem now.
\begin{enumerate}
\item[(a)] Let us suppose $\limsup\limits_n\frac{\ln P_{n+1}}{\ln P_n}=\beta-1$.
\begin{enumerate}
\item[(a1)] We will prove that
$$i\notin \bigcup\limits_{\delta>0}J_{\beta+\delta}.$$
Combining Eqs.(\ref{claim2}) and (\ref{app1}) we have
$$\left|i-\frac{Q_n}{P_n}\right|>\frac{1}{P_n^{\beta+\delta}},$$
$n\geq n_{\delta}$ for the $\delta$ and the $n_{\delta}$ in {\bf Claim 2}.

This inequality, together with {\bf Claim 1}, imply that the set
$$\left\{\frac rs\in\Bbb Q: \;\left|i-\frac rs\right|<\frac 1{s^{\beta+\delta}}\right\}$$
has finite cardinality, which implies 
$i\notin J_{\beta+\delta}$, for any $\delta>0$, which is (a1). 
\item[(a2)] We will prove:
$$i\in\bigcap\limits_{\beta-2\geq\theta>0}J_{\beta-\theta}.$$
Combining Eq.(\ref{claim3}) and the right part of Eq.(\ref{app1}) we have that, if $i\in (0,1)$, there exists a sequence $n_j=n_j(\theta)$ of natural numbers such that
$$\left|i-\frac{Q_{n_j}}{P_{n_j}}\right|<\frac{1}{P_{n_j}^{\beta-\theta}},$$
which implies $i\in J_{\beta-\theta}$
which, in turn, implies (a2).
\end{enumerate}
\item[(b)]Let us suppose
$$i\in\bigcap\limits_{\beta-2\geq\theta>0}J_{\beta-\theta}- \bigcup\limits_{\delta>0}J_{\beta+\delta}.$$
We will prove that $\limsup\limits_n\frac{\ln P_{n+1}}{\ln P_n}=\beta-1$.

Since $i\notin\bigcup\limits_{\delta>0}J_{\beta+\delta}$, then for each $\delta>0$ the inequality
$\left|i-\frac{Q_n}{P_n}\right|<\frac 1{P_n^{\beta+\delta}}$
has finite solutions.
So 
$\exists n(\delta)\in\Bbb N$ such that
$$\left|i-\frac{Q_n}{P_n}\right|\geq\frac 1{P_n^{\beta+\delta}} \;\forall\;n\geq n(\delta).$$
Also, the right side of Eq.(\ref{app1}) implies
$P_nP_{n+1}<P_n^{\beta+\delta}$,
that is
$P_{n+1}<P_n^{\beta+\delta-1}$.
Therefore $\forall n\geq n(\delta)$
we have
$\frac{\ln P_{n+1}}{\ln P_n}<\beta+\delta-1,$
hence
$$\limsup\limits_{n\to\infty}\frac{\ln P_{n+1}}{\ln P_n}\leq \beta+\delta-1.$$
This inequality holds $\forall\delta>0$, therefore  
$$\limsup\limits_{n\to\infty}\frac{\ln P_{n+1}}{\ln P_n}\leq \beta-1.$$
Now, if we had
$$\limsup\limits_{n\to\infty}\frac{\ln P_{n+1}}{\ln P_n}< \beta-1,$$ then, there would exist $\theta_0>0$ such that 
$$\limsup\limits_{n\to\infty}\frac{\ln P_{n+1}}{\ln P_n}= \beta-\theta_0-1,$$
and from part (a) above we would have:
$$i\notin\bigcup\limits_{\delta>0}J_{\beta-\theta_0+\delta},$$
from which $i\notin J_{\beta-\theta}$ for any $\theta>0$ small enough,
which is a contradiction with
$$i\in\bigcap\limits_{0<\theta\leq \beta-2}J_{\beta-\theta},$$
a part of our hypothesis.
\end{enumerate}

It remains to prove the three Claims.

{\bf Proof of Claim 1.}
Let $\frac rs$ be a rational number fulfilling 
$$\left|i-\frac rs\right|<\frac{1}{s^{\beta+\theta}},$$
and such that
$s\geq [2^\frac 1{\theta}]+1$.
This inequality, in turn, implies
$s^{\beta+\theta}>2s^{\beta}$, and since $\beta\geq 2$, $s\in\Bbb N$, we have $s^{\beta}\geq s^2$. Therefore
$s^{\beta+\theta}>2s^2.$
From this and
$$\left|i-\frac rs\right|<\frac{1}{s^{\beta+\theta}}$$
we have
$$\left|i-\frac rs\right|<\frac 1{2s^2}.$$
From this and {\bf Property 1} we have
$\frac rs=\frac{Q_n}{P_n}$ for some $n\in\Bbb N$, q.e.d.

Before proving {\bf Claim 2} we need an

{\bf Observation.} 1), 2) and 3) below are equivalent statements.
\begin{enumerate}
\item 
$\limsup\limits_{n\to\infty}\frac{\ln P_{n+1}}{\ln P_n}=\beta-1$
\item
\begin{enumerate}
\item 
$\forall\theta>0,\;\exists n(\theta)\in\Bbb N\;\mbox{such that}\;\frac{\ln P_{n+1}}{\ln {P_n}}<\beta-1+\theta\;\forall\;n\geq n(\theta).$
\item
$\forall\theta>0,\;\exists\left\{n_j(\theta)\right\}$, an infinite sequence of natural numbers, such that
$$\beta-1-\theta<\frac{ln P_{n_j(\theta)+1}}{\ln P_{n_j(\theta)}}.$$
\end{enumerate}
\item
\begin{enumerate}
\item[(a)] 
$\forall\theta>0,\;\exists n(\theta)\in\Bbb N\;\mbox{such that}\; P_{n+1}<P_n^{\beta-1+\theta}\;\forall\;n\geq n(\theta).$
\item[(b)] 
$\forall\theta>0,\;\exists\left\{n_j(\theta)\right\}$, an infinite sequence of natural numbers, such that $$P_{n_j(\theta)}^{\beta-1-\theta}< P_{n_j(\theta)+1}.$$
\end{enumerate}
\end{enumerate}

{\bf Proof of Claim 2.} Let $\theta$ and $\delta$ be positive numbers, $\theta<\delta$. 
Our hypothesis implies that $\exists n(\theta)\in\Bbb N$ such that
$$P_{n+1}<P_n^{\beta-1+\theta}$$
$\forall n\geq n(\theta)$. From this we have
\begin{equation}
\label{aux2}
P_n(P_n+P_{n+1})<P_n^2+P_n^{\beta+\theta}=P_n^{\beta+\theta}\left(P_n^{2-\beta-\theta}+1\right), 
\end{equation}
and since
$$\frac{P_n^{\beta+\theta}\left(P_n^{2-\beta-\theta}+1\right)}{P_n^{\beta+\delta}}=P_n^{\theta-\delta}\left(P_n^{2-\beta-\theta}+1\right)$$
and
$$\lim\limits_{n\to\infty}P_n^{\theta-\delta}\left(P_n^{2-\beta-\theta}+1\right)= 0$$
we have that $\exists n_0\in\Bbb N$ such that, $\forall n>n_0$ 
\begin{equation}
\label{aux5}  
P_n^{\beta+\theta}\left(P_n^{2-\beta-\theta}+1\right)<P_n^{\beta+\delta}
\end{equation} 
holds.

Now, if $n>n(\delta)=\max \{n_0, n(\theta) \}$, Eqs. (\ref{aux2}) and (\ref{aux5}) hold simultaneously, from which $$P_n(P_n+P_{n+1})<P_n^{\beta+\delta}\;\forall n\geq n(\delta),\;\;\mbox{q.e.d.}$$

{\bf Claim 3} is an obvious consequence of statement 3)(b) in the Observation above.

{\bf Theorem 2}

With the notation in Theorem 1 we have
\begin{enumerate}
\item[(a)] 
$\bigcap\limits_{0<\theta\leq \beta-2}J_{\beta-\theta}\neq J_{\beta} $
\item[(b)] 
$\bigcup\limits_{\delta>0}J_{\beta+\delta}\neq J_{\beta}$
\end{enumerate}

{\bf Proof.} With bits and pieces in the proof of Theorem 1, a proof of (a) and (b) can be put together, a technical exercise we leave to the reader.

\section{The Multifractal Spectrum of $\Omega$}
\subsection{The Cantordust Set}
The fractal set $\Omega$ underlying the Cantor staircase $q=g(-H)$ studied by Bruinsma and Bak can be constructed in a way analogous to the one used  to construct the Cantordust sets in section 1.3. Let $\Delta H(\frac{Q}{P})$ be the resonance interval corresponding to the rational number $\frac{Q}{P}$, that is the step $\Delta H(\frac{Q}{P})$ in the staircase of height $\frac{Q}{P}$. We subtract from the real line $\Bbb R$ intervals $\Delta H(\frac{0}{1})$ and $\Delta H(\frac{1}{1})$, obtaining a closed and bounded interval $I^0$; this will be the equivalent of the initial interval $I=[0,1]$ in the two examples in section 1.3. Next, we subtract interval $\Delta H(\frac{1}{2})$ from $I^0$, thereby obtaining two compact intervals $I^1_1$ and $I^1_2$. Intervals $I^1$ constitute the first approximation to $\Omega$. Next, we subtract $\Delta H(\frac{1}{3})$ and $\Delta H(\frac{2}{3})$ from intervals $I^1$ obtaining four compact intervals $I^2_i,\;i=1,...,4$ which constitute the second approximation to $\Omega$. Proceeding in this way, in the $k$-th step we subtract all intervals $\Delta H(\frac{Q}{P}),\;\frac{Q}{P}$ in the $(F-B)_k$ sequence, thereby obtaining $2^k$ intervals $I^k_i,\;i=1,...,2^k$; which are the $k$-th approximation to $\Omega$.
\subsection{The Measure of Probability on $\Omega$} 
The probability $p$ induced in our Cantordust $\Omega$ will be like in section 1.3., given by $p_1=p(I^1_1\cap \Omega)=\frac12$, $p_2=p(I^1_2\cap \Omega)=\frac12$. All segments $I^k$ will be equiprobable, i.e.  $p(I^k_i\cap \Omega)=\frac{1}{2^k}$ for $i=1,...,2^k$.

Notice that $g(I^1_1\cap\Omega)=[0,\frac{1}{2}]$ and  $g(I^1_2\cap\Omega)=[\frac{1}{2},1]$, whereas $g(I^2_1\cap\Omega)=[0,\frac{1}{3}]$, $g(I^2_2\cap\Omega)=[\frac{1}{3},\frac{1}{2}]$, $g(I^2_3\cap\Omega)=[\frac{1}{2},\frac{2}{3}]$, and $g(I^2_4\cap\Omega)=[\frac{2}{3},1]$.

Therefore, the equiprobability of the $I^k$ approximating $\Omega$ is equivalent to the equiprobability of the $2^k$ segments in the $(F-B)_k$ partition. Such equiprobability on the $(F-B)_k$ segments induces a probability measure in the unit segment --the hyperbolic measure-- a measure that we will study below.

\subsection{$\alpha_{\max}$ of $\Omega$}
Let us go back to section 1.3. Let $x\in \Omega$. Let $I^k=I^k(x),\; k\in\Bbb N$, be the sequence of nested intervals in successive $k$-approximations to $\Omega$ to which $x$ belongs. Let us recall that the equation for the $\alpha$-index of $I^k$ is $p(I^k\cap \Omega)=\left|I^k\right|^{\alpha}$. This sequence $\alpha=\alpha(I^k)=\alpha_k,\;k\in\Bbb N$, when convergent, defines $\alpha(x)$. Notice that $p(I^k\cap \Omega)=\frac{1}{2^k}$, therefore $\alpha_k$ depends strictly on the size $\left|I^k\right|=\left|I^k(x)\right|$. Hence, if we are interested in points $x\in\Omega$ with $\alpha(x)=\alpha_{\max}$ we have to select points for which intervals $I^k(x)$ are the longest in the $k$-approximation to $\Omega,\;k\in\Bbb N$.

Let us consider the point in the staircase of height $g(x)=i$, an irrational value. The steps near it were called $\Delta H_{i,\varepsilon}$ and we will call them $\Delta H(i)$ for short. We have $\Delta H(i)=\Delta H(g(x))$. We constructed $I^k(x)$ by subtracting intervals $\Delta H(i)=\Delta H(g(x))$. Therefore {\bf long} $I^k(x)$ correspond to {\bf small} $\Delta H(g(x))$, and viceversa. We are interested now in {\bf small} $\Delta H(i)$.

Small $\Delta H(i)$ correspond to steps near points $i\in G_2$. Among them, the smallest correspond to points $i=[a_1,...,a_n,1,...,1,...]$ and the smallest of them all correspond to $$i=[1,1,...,1,...]=\phi^{-1}=\phi-1,$$ $\phi$ being the golden mean, a result that agrees with classical ones. 

Notice that in the vertical $q$-axis provided with the $(F-B)$ arrangement, the {\bf short} segments in each $(F-B)_k$ partition are precisely the ones that cover the point $i=\phi^{-1}=\phi-1$. Therefore, when $I=[0,1]$ is provided with the Hyperbolic measure of probability, we have that points in $I$ with $\alpha_{\min}$ correspond to points $x$ in $\Omega$, $i=g(x)$, with $\alpha_{\max}$.

\subsection{$\alpha_{\min}$ of $\Omega$}
Following the outlines in the preceding section, points $x$ with $\alpha_{\min}$ in $\Omega$ are those with the shortest $\left|I^k(x)\right|$, which in turn correspond to the longest $\Delta H(g(x))=\Delta H(i)$. The longest $\Delta H(i)$ correspond to $i\in G_{\infty}=J_{\infty}$ [Piacquadio and Grynberg, 1998].

{\bf Liouville Numbers.} Liouville constructed irrational numbers $i$ for which, for each $k\in\Bbb N$, there exists rationals $\frac QP$ such that
$$\left|i-\frac QP\right|<\frac{1}{P^k},\;\;P\geq 2.$$
Let $a_1$ be arbitrary. Choose $a_2>P_1^1(a_1)$ and, given $P_{n-1}(a_1,...,a_{n-1}),$ we choose $a_n>P_{n-1}^{n-1}(a_1,...,a_{n-1})$.
Such growth of the $a_n$ guarantees (see properties of $P_n$ in Sec.1.4 a)) that $i=[a_1,a_2,...,a_n,...]$ belongs to every $J_k$, hence, to $J_{\infty}=G_{\infty}$. Notice that these $a_n$ have an ultrarapid growth, that $a_{n+1}>>a_n\;\forall\;n\in\Bbb N$. Therefore, the part of the continued fraction expansion of $i$ that starts with $a_n$, i.e. $a_n+\frac{1}{a_{n+1}+\frac{1}{\ddots}}$ is almost indistinguishable from $a_n$.

Let us recall that a real number is a rational number precisely when its continued fraction has a finite number of such $a_n$. Therefore, the elements of $G_{\infty}$ are called time and again "quasi-rationals" in the literature.

Therefore we will study rational numbers in the vertical axis, because they  have properties analogous to those of "quasi-rational" numbers, i.e. we will study quasi-rationals in $G_\infty$ via rational numbers.

Let $\frac{Q}{P}$ and $\frac{Q'}{P'}$ be two rationals adjacent in a certain $(F-B)_k$. In order to avoid overlapping we will consider segments in an $(F-B)_k$ partition as closed on the left and open on the right: $[\frac{Q}{P},\frac{Q'}{P'})$ will be our $(F-B)_k$ segment covering point $\frac{Q}{P}$. The only segment in $(F-B)_{k+1}$ covering $\frac{Q}{P}$ is $[\frac{Q}{P},\frac{Q+Q'}{P+P'})$; the one in $(F-B)_{k+2}$ is $[\frac{Q}{P},\frac{2Q+Q'}{2P+P'})$, and in general, in $(F-B)_{k+n}$ it will be $[\frac{Q}{P},\frac{nQ+Q'}{nP+P'})$. The length of such general segment is $\frac{1}{P(nP+P')}\cong\frac{1}{P^2}\frac{1}{n}$, a value that diminishes like the harmonic sequence.

So, while the harmonic sequence $\frac{1}{n}$ is responsible for the longest segments, the sequence $\frac{1}{\phi^{2n}}$ is responsible for the shortest ones.

\subsection{The Value $\alpha\in(\alpha_{\min},\alpha_{\max})$ and the Type $G_{\beta}$}

Let us recall that $\Omega_{\alpha}\subset\Omega$ is the set of all points $x\in\Omega$ that share the same $\alpha$-concentration, i.e. $\Omega_{\alpha}=\left\{x\in\Omega:\;\alpha(x)=\alpha\right\}$. Now, as we remarked in Secs. 4.3 and 4.4, $\alpha(x)$ is large (small) if the nested $I^k(x)$, $k\in\Bbb N$ (from successive $k$-partitions of $\Omega$), are large (small).

We also remarked (same Secs.) that the size of $I^k(x)$ was related to the size of $\Delta H(i)=\Delta H(g(x))$: the larger $\Delta H(g(x))$, the smaller $I^k(x)$ --hence $\alpha(x)$-- will be. But the size of steps $\Delta H(i)=\Delta H(g(x))$ with height $\cong i$ in the staircase, depends strongly on the $G_{\beta}$ to which $i=g(x)$ belongs. We are saying that if $i_1\in G_{\beta}$ and $i_2\in G_{\beta}$ for the same $\beta$, then $x_1=g^{-1}(i_1)$ and $x_2=g^{-1}(i_2)$ should belong to the same $\Omega_{\alpha}$, for some $\alpha=\alpha(\beta)$. In other words: given $\beta\geq 2$ there should be $\alpha=\alpha(\beta)$ such that $g^{-1}(G_{\beta})=\Omega_{\alpha(\beta)}$. While  we cannot yet categorically affirm the validity of this equality, the remarks above show that the $G_{\beta}$ and the $\Omega_{\alpha}$ are closely linked.

\subsection{The Spectrum $(\alpha,f(\alpha))$ of $\Omega$}
Let us consider $\alpha$ growing from $\alpha_{\min}$ to $\alpha_{\max}$, and let us for a moment accept that $g(\Omega_{\alpha})=g(\Omega_{\alpha(\beta)})=G_{\beta}$. Then we have $G_{\beta}$ changing from $G_{\infty}$ to $G_2$ as $\alpha$ grows. Notice that $f(\alpha)=d_H(\Omega_{\alpha})=d_H(\Omega_{\alpha(\beta)})$ would be directly related --via the {\bf increasing} function $g$-- to $d_H(G_{\beta})$, which strictly grows when $\beta$ changes from $\infty$ to $2$, going from $d_H(G_{\infty})=0$ to $d_H(G_{2})=1$. The function $g$ linking $\Omega_{\alpha(\beta)}$ and $G_{\beta}$, and the uneven (and hitherto poorly understood) distribution of intervals $I^k$ in a $k$-partition of $\Omega$, are two strong factors which hinder us from linking $f(\alpha)=d_H(\Omega_{\alpha(\beta)})$ with $d_H(G_{\beta})$ directly through, say, so simple a way as an equality. Still, we know:
\begin{enumerate}
\item
For $\beta=\infty$ we have $\alpha(\beta)=\alpha_{\min}$.
\item
For $\beta=2$ we have $\alpha(\beta)=\alpha_{\max}$. 
\item
$f(\alpha_{\min})=0$.
\item
$f(\alpha)$ is increasing for the greater part of the interval $[\alpha_{\min},\alpha_{\max}]$.
\item
Let $\max\limits_{\alpha}f(\alpha)=f(\alpha^{\max})$. Then $\alpha_{\max}$ is very near $\alpha^{\max}$.
\item
The $\Omega_{\alpha}$ are strongly linked to the types $G_{\beta}$.
\item
The sets $G_{\infty}$ and $G_{2}$, related to $\alpha_{\min}$ and $\alpha_{\max}$ in $\Omega$, are related to $\alpha_{\max}$ and $\alpha_{\min}$ in $I=[0,1]$ endowed with the hyperbolic measure of probability induced by Farey-Brocot.
\end{enumerate}

\section{The Spectrum $(\alpha, f(\alpha))$ of the Fractal Set $\Omega$ underlying the Circle Map Staircase}

Conclusions 1) to 7) in the last section show a strong connection between the magnetization function $q=g(-H)$ and leading problems in Number Theory --viz the good approximation of irrational numbers studied with Jarn\'{\i}k classes $J_{\beta}$ and their refinements $G_{\beta},\;\beta\geq 2$. This particular connection between magnetization and Number Theory is seen only when analyzing the multifractal spectrum of the fractal set $\Omega$ underlying the magnetization Cantor staircase. The conclusions (1) to 7)) are based, as we have seen,  on two premises about the staircase $q=g(-H)$: the $(F-B)$ arrangement of the stairsteps $\Delta H$ in the staircase, and the formula given by Eq. (\ref{fases}).

Let us now consider the Cantor staircase $W=g(\omega)$ associated with the circle map: we land in Dynamical Systems, where connections with Number Theory are old and well explored. The stairsteps $\Delta\omega$ do satisfy the $(F-B)$ arrangement [Cvitanovic et al., 1985], as we remarked above. On the other hand, Fig. 1 shows that Eq. (\ref{fases}) is valid with $\Delta\omega(\frac QP)$ instead of $\Delta H(\frac QP)$ --at least when we take averages over the $\Delta\omega(\frac QP)$ with the same $P$.

We can, therefore, extend conclusions 1) to 7) for the case of the circle map staircase $W=g(\omega)$. Now, for the fractal set $\Omega$ underlying staircase $W=g(\omega)$ there is a spectrum $(\alpha, f(\alpha))$ associated with it [Halsey et al., 1986]. A natural question arises: are conclusions 1) to 7) verifiable for this $f(\alpha)$?

Conclusions 1) and 2) can be checked from statements as early as "...the most extremal behaviours of this staircase are found around the golden mean sequence of dressed winding numbers... and at the harmonic sequence $\frac{1}{Q}\to 0$. The most rarified region of the staircase is located around the golden mean" "...the $\frac{1}{Q}$ series... determines the most concentrated portion of the staircase..." in [Halsey et al.,1986]. Conclusions 3) and 5) follow from observing the corresponding graph $(\alpha, f(\alpha))$ in Fig. 12, in the same reference. Conclusion 4) follows observing the same figure: $f(\alpha)$ is increasing for an interval $\Delta\alpha\subset[\alpha_{\min},\alpha_{\max}]$, where the length $|\Delta\alpha|$ is some $98\%$ of the length $|[\alpha_{\min},\alpha_{\max}]|$. Conclusion 7) holds just as it does for $q=g(-H)$. Conclusion 6) remains a qualitative one --and quantitatively conjectural.

If we ponder on the fact that the Circle Map is universal in character, i.e. with small change of details, it describes a variety of phenomena, and if we recall that the time variables $W$ and $\omega$ in the function $W=g(\omega)$ have no connection with $q$ and the magnetic field $H$, then we can safely conclude that Number Theory links with Cantor staircases in physics in a way even more universal than the one indicated in the already explored linking with Circle Maps and Dynamical Systems.

\section{The Hyperbolic Metric}

Let $\Bbb H=\left\{z\in\Bbb C:\;Im(z)>0\right\}$ be the upper half plane. The geodesics are circumferences orthogonal to the real axis, and that includes semilines orthogonal to the real axis. The congruences are transformations
$$z\to\frac{az+b}{cz+d},\;a,b,c,d\; \mbox{in}\; \Bbb Z,\;ad-bc=1$$
The congruences have a group structure denoted by ${\cal U}$ in the literature. We can consider ${\cal U}$ as a multiplicative group of $2\times 2$ matrices 
$\left(\begin{array}{cc}
a&b\\
c&d\\
\end{array}\right)$
with integer entries and unit determinant.

Associated with ${\cal U}$ there is a fundamental region $R\subset\Bbb H$ such that if $z\in Int R$ then $uz\in Ext R$ for every $u\in {\cal U}$, $u\neq Identity$; and given $z\in Ext R$ there exists $u\in {\cal U}$ such that  $uz\in Int R$. Points on the boundary of $R$ are transformed, by some elements of ${\cal U}$, into other points on the same boundary. All regions $uR$, $u\in {\cal U}$, have disjoint interiors. The union of all $uR$, $u\in {\cal U}$, covers $\Bbb H$. Such $R$ is called a fundamental tile. Any $uR$ is another fundamental tile. Any two such tiles are congruent by means of an element in ${\cal U}$. Looking at the tiling we notice that tiles near the real axis are much smaller than other tiles. Yet, if we take off our Euclidean eyeglasses, put on a pair of Hyperbolic spectacles, and look again at the tiling, we will see all tiles equal to one another very much like, say, squares of the same size: for there exists a unique metric --the hyperbolic one-- measuring with which the sizes of all tiles are the same.

Matrices 
$P=\left(\begin{array}{cc}
1&1\\
0&1\\
\end{array}\right)$
and
$Q=\left(\begin{array}{cc}
0&1\\
-1&0\\
\end{array}\right)$
generate ${\cal U}$, $R$ being the fundamental tile situated symmetrically above the origen.
We take fundamental tile $R$ and perform on it the cut-and-paste surgery indicated in [Series, 1985; Grynberg and Piaquadio, 1995] obtaining another fundamental tile $T$ for ${\cal U}$. The generators of ${\cal U}$ are now matrices $P$ and $A=QP^{-1}Q=\left(\begin{array}{cc}
1&0\\
1&1\\
\end{array}\right)$, that is, given $u\in {\cal U}$ there exists a finite word in letters $A$ and $P$ such that $u=A^{a_1}P^{a_2}A^{a_3}P^{a_4}\dots\;a_i\in\Bbb N$. We denote tile $uT=A^{a_1}P^{a_2}A^{a_3}P^{a_4}\dots T$ by the finite word $A^{a_1}P^{a_2}A^{a_3}P^{a_4}\dots$ Now, $T$ is a "rhombus", and so is tile $A$. Two opposite vertices of rhombus $A$ are points $0$ and $1$ of the real axis: we associate $A$ with segment $[0,1]$. Next, we will consider tiles associated with words starting with letter $A$, and such that all $a_i\in\Bbb N$. Two-letter words like that are $AA$ and $AP$. $AA$ is the rhombus associated with segment $[0,\frac 12]$: it is the only tile with two opposite vertices leaning on $0$ and $\frac12$ on the real line. $AP$ is likewise associated with $[\frac12,1]$. Tiles $AA$ and $AP$ are Euclideanly smaller than $A$ and are closer to $\Bbb R$ than is $A$. Three-letter words $A^3,A^2P,APA$ and $AP^2$, associated with intervals $[0,\frac13],[\frac13,\frac12],[\frac12,\frac23]$ and $[\frac23,1]$, respectively, are Euclideanly smaller than two-letter tiles, and are even closer to $\Bbb R$. Letter $A$, therefore, is associated with the "left", and $P$ to the "right", in a way we trust is obvious. The $2^k$ words with $k$-letters are associated with the $(F-B)_k$ sequence.

An infinite word $A^{a_1}P^{a_2}A^{a_3}P^{a_4}\dots$, therefore, is associated with an irrational number $i$ in the unit segment; moreover

$$A^{a_1}P^{a_2}A^{a_3}P^{a_4}\dots=[a_1,a_2,a_3,a_4,...]=\frac{1}{a_1+\frac{1}{a_2+\frac{1}{a_3+\frac{1}{\ddots}}}}=i.$$

With $C$ indicating $A$ or $P$, according to the case, we have that tile 
$$A^{a_1}P^{a_2}A^{a_3}P^{a_4}\dots C^{a_n}$$
is a $2\times 2$ matrix with unit determinant
$$\left(\begin{array}{cc}
Q_{n-1}&Q_n\\
P_{n-1}&P_n\\
\end{array}\right)$$
where $\frac{Q_k}{P_k}$ is $[a_1,a_2,...,a_k]$.

Tiles $A^{k+1},...,AP^k$ corresponding to the $2^k$ words starting with $A$ followed by $k$ letters $A$ and $P$, are hyperbolically equimeasurable, and they are associated with the $2^k$ segments in $(F-B)_k$, for all $k\in\Bbb N$. Then we say that $[0,1]$ inherits from $\Bbb H$ a measure $\mu$ that renders these $2^k$ segments equimeasurable, the $\mu$ measure of each segment being $\frac{1}{2^k}$. {\it Par abus de langage} we will refer to this measure $\mu$ in $[0,1]$ indistinctly as the hyperbolic measure or the $(F-B)$ measure. 

\section{Hyperbolic Self-similarity of the Cantor Staircase}
Let $y=g(x)$ be any Cantor staircase that fulfills both Eq. (\ref{fases}) for the intervals of resonance and the $(F-B)$ arrangement for the vertical axis. The Circle map and the magnetization curve both fulfill this condition. 

It has been claimed [Bruinsma and Bak, 1983; Bak, 1986] that the graph of $y=g(x)$ is "self-similar": the whole staircase looks like a small section of it. In this section we study the character of this self-similarity: we will focus on a section of the staircase, say, the section between two small intervals of resonance $I$ and $I'$. {\it Par abus de langage}, with $I$ and $I'$ we will denote the corresponding stairsteps as well. For clarity we will choose $I$ and $I'$ with heights $\frac{Q}{P}$ and $\frac{Q'}{P'}$ in the staircase, where $\frac{Q}{P}$ and $\frac{Q'}{P'}$ are two rational numbers adjacent in a $(F-B)_k$ partition, for a certain $k\in\Bbb N$.

Let us consider the whole staircase, situated between steps $I(\frac{0}{1})$ and $I(\frac{1}{1})$, see Sec. 2. In the staircase there is a specific way in which the height of intervals of resonance is distributed according to the Euclidean size of the latter. Such relationship between height and size is precisely what we described as the $(F-B)$ arrangement of the vertical axis of the  staircase. Let us consider now the staircase between $I=I(\frac{Q}{P})$ and $I'=I'(\frac{Q'}{P'})$. The segment $[\frac{Q}{P}, \frac{Q'}{P'}]$ is obtained from $[\frac01,\frac11]$ by means of a hyperbolic rigid movement. Such hyperbolic movement is given by a $(k+1)$-letter word in letters $A$ and $P$, starting with $A$. Therefore the distribution of heights of steps in the staircase between $I$ and $I'$ is hyperbolically equivalent to that of the whole staircase.

Let ${\cal T}={\cal T}(b_1,b_2,...,b_m)=A^{b_1}P^{b_2}\dots C^{b_m}$, $b_1+b_2+\dots+b_m=k+1$ be the corresponding $(k+1)$-letter word effecting the hyperbolic transformation, and let $i=[a_1,a_2,...,a_n,...]$ be the irrationals in $[0,1]$. Then the irrationals in $[\frac{Q}{P}, \frac{Q'}{P'}]$ are written $[b_1,...,b_m;a_1,a_2,...,a_n,...]$.

Let us focus in this general expression of an irrational in $[\frac{Q}{P}, \frac{Q'}{P'}]$: the first $m$-numbers $b_1,...,b_m$ give the hyperbolic transformation 
$${\cal T}=A^{b_1}P^{b_2}\dots C^{b_m},$$ the numbers that follow, $a_1,a_2,...,a_n,...$ give all the irrationals in $[0,1]$. Reading from left to right, this notation $[b_1,...,b_m;a_1,a_2,...,a_n,...]$ given by continued fractions for an irrational in $[\frac{Q}{P}, \frac{Q'}{P'}]$ yields this number in a natural way as ${\cal T}=A^{b_1}P^{b_2}\dots C^{b_m}$ applied to the irrationals $[a_1,a_2,...,a_n,...]$ of $[0,1]$.

We have seen, then, that heights of stairsteps vis-$\grave{a}$-vis their size is, hyperbolically, the same for steps between $\frac{Q}{P}$ and $\frac{Q'}{P'}$ and for stairsteps between $\frac01$ and $\frac11$. Next, we have to compare sizes of steps between $\frac{Q}{P}$ and $\frac{Q'}{P'}$ with the size of the corresponding steps between $\frac01$ and $\frac11$.

\subsection{The Change of Scale for the Stairsteps}
Let ${\cal T}$ be the hyperbolic transformation just described: 
$${\cal T}[a_1,a_2,...,a_n,...]=[b_1,b_2,...,b_m;a_1,a_2,...,a_n,...],$$
${\cal T}:[0,1]\to [\frac{Q}{P},\frac{Q'}{P'}].$

Intervals of stability with height $[a_1]$,$[a_1,a_2]$,...,$[a_1,a_2,...,a_n]$,... get nearer and nearer the point in the staircase of height $[a_1,a_2,...,a_n,...]$. The heights of the associated steps in the staircase between $I(\frac{Q}{P})$ and $I'(\frac{Q'}{P'})$ are $$[b_1,b_2,...,b_m;a_1],\;[b_1,b_2,...,b_m;a_1,a_2],...,\;[b_1,b_2,...,b_m;a_1,a_2,...,a_n],...$$

Let us recall that $[a_1,...,a_n]=\frac{Q_n}{P_n}$ is a good approximant of irrational $[a_1,...,a_n,...]$. $P_n$ is a polynomial $P_n(a_1,...,a_n)$. Let us also recall that the size of $I(\frac{Q}{P})$ is given by $\frac{1}{P}$. Therefore, lengths of steps $[a_1,a_2,...,a_n]$ and the associated steps $[b_1,b_2,...,b_m;a_1,a_2,...,a_n]$ are, respectively, given by
$$\frac{1}{P_n(a_1,...,a_n)}$$
and
$$\frac{1}{P_{m+n}(b_1,b_2,...,b_m;a_1,...,a_n)}.$$
How does the size of these two intervals compare? The corresponding scale factor $\lambda$, if it exists, would be
$$\lambda= \frac{\frac{1}{P_{m+n}(b_1,b_2,...,b_m;a_1,...,a_n)}}{\frac{1}{P_n(a_1,...,a_n)}}=\frac{P_n(a_1,...,a_n)}{P_{m+n}(b_1,b_2,...,b_m;a_1,...,a_n)}=$$
$$\frac{P_n(a_1,...,a_n)}{P_m(b_1,...,b_m)P_n(a_1,a_2,...,a_n)+P_{m-1}(b_1,...,b_{m-1})P_{n-1}(a_2,...,a_n)}.$$
In order to see this the reader is invited to decompose, say,
$$P_1(b_1),\;P_2(b_1,b_2),\;P_3(b_1,b_2,a_1),\;P_4(b_1,b_2,a_1,a_2)...$$ and verify that the decomposition of these $P's$ in terms of $P(b's\;\mbox{only:}\;b_1, b_2)$ and $P(a's\;\mbox{only: as many}\; a's\; \mbox{as you can handle})$ is
$$P_4(b_1,b_2,a_1,a_2)=P_2(b_1,b_2)P_2(a_1,a_2)+P_1(b_1)P_1(a_2).$$
With {\bf many more lines!} for the general expresion we can see that
$$P_{m+n}(b_1,b_2,...,b_m,a_1,a_2,...,a_n)=$$
$$P_m(b_1,b_2,...,b_m)P_n(a_1,a_2,...,a_n)+
P_{m-1}(b_1,b_2,...,b_{m-1})P_{n-1}(a_2,...,a_n)$$
holds. Therefore,
$$\lambda=\frac{P_n(a_1,...,a_n)}{P_m(b_1,...,b_m)P_n(a_1,a_2,...,a_n)+P_{m-1}(b_1,...,b_{m-1})P_{n-1}(a_2,...,a_n)}=$$
$$\frac{P_n(a_1,...,a_n)}{P_m(b_1,...,b_m)P_n(a_1,...,a_n)\left(1+\theta_m\right)}=$$
$$\frac{1}{P_m(b_1,...,b_m)\left(1+\theta_m\right)}=\left(\frac{1}{1+\theta_m}\right)\frac{1}{P_m(b_1,...,b_m)},$$ 
where
$$\theta_m=\frac{P_{m-1}(b_1,...,b_{m-1})}{P_m(b_1,...,b_m)}\frac{P_{n-1}(a_2,...,a_n)}{P_n(a_1,a_2,...,a_n)}.$$
Since $$P_{m-1}(b_1,...,b_{m-1})<P_m(b_1,...,b_m)$$
and
$$P_{n-1}(a_2,...,a_n)<P_n(a_1,a_2,...,a_n)$$
then $\theta\in (0,1)$ and $\frac{1}{1+\theta}\in (\frac12,1)$, hence
$$\frac12\frac{1}{P_m(b_1,...,b_m)}<\lambda<\frac{1}{P_m(b_1,...,b_m)},$$
i.e. $\lambda$ is of the order of $\frac{1}{P_m(b_1,...,b_m)}$. Moreover, remarks in Sec. 1.4 on the growth of polynomials $P_n$ and the number of monomials in $P_n$ imply that the quotient of two successive $P_n$ and $P_{n-1}$ is no smaller than $\phi$, save for some pathological (and enumerable) cases. Hence our $\theta$ fluctuates between $0$ and $\frac{1}{\phi^2}$ and
$$0,723...\frac{1}{P_m(b_1,...,b_m)}=\frac{1}{1+\frac{1}{\phi^2}}\frac{1}{P_m(b_1,...,b_m)}<\lambda<\frac{1}{P_m(b_1,...,b_m)}.$$

{\bf A comment.} Let us have a look at $\theta$: we have in $\theta$ the product of two quotients of consecutive polynomials $P$. Let us consider two such consecutive polynomials $P_{n-1},\;P_n$ in variables $j_1,j_2,...,j_n,...,\;j_i\in\Bbb N$. Roughly speaking, these variables, being natural, can either grow or not. If they grow, we have $\frac{P_{n-1}}{P_n}\to 0$. If they don't grow, the most extreme case is $1,1,...,1,...,$ where $\frac{P_n}{P_{n-1}}\cong\phi$. That is why $\frac{P_{n-1}}{P_n}$ fluctuates between $0$ and $\frac{1}{\phi}$. Hence our bounds on $\theta$.

We have, then, that the Cantor staircase is hyperbolically self-similar in the vertical axis, whereas the size of horizontal stairsteps have an Euclidean self-similar structure: sizes of steps of height between $\frac{0}{1}$ and $\frac{1}{1}$ change to sizes of steps of height between $\frac{Q}{P}$ and $\frac{Q'}{P'}$, with a contractor  
$$\lambda\cong\frac{1}{P_m(b_1,...,b_m)},$$
where $b_1,...,b_m$ defines the hyperbolic transformation ${\cal T}:[0,1]\to [\frac{Q}{P},\frac{Q'}{P'}].$

There are two modes of self-similarity involved here: one hyperbolic and one Euclidean. Vertical segments $[\frac{Q}{P},\frac{Q'}{P'}]$ have length
$$\frac{1}{PP'}=\frac{1}{P_m(b_1,...,b_m)P_{m-1}(b_1,...,b_{m-1})},$$
whereas the size of horizontal segments between $\frac{Q}{P}$ and $\frac{Q'}{P'}$ decrease with a factor $\lambda\cong\frac{1}{P_m(b_1,...,b_m)}$:
hence vertical sizes and horizontal sizes decrease with diferent scale factors.

Nevertheless, there is a particular sense in which the size of horizontal segments (stairsteps) also decreases according to a hyperbolic law: we explained above that vertical segment $[\frac{Q}{P},\frac{Q'}{P'}]$ is $[\frac{Q_n}{P_n},\frac{Q_{n-1}}{P_{n-1}}]$. Now, we also stated that rationals $\frac{Q_n}{P_n}$ well approximated a certain irrational $i$. We also stated that the size of stairsteps $I(\frac{Q_n}{P_n})$ near $i$ behaves according to the type $G_\beta$ to which $i$ belongs. By "near $i$" we mean [Piacquadio and Grynberg, 1998] steps situated exactly between $\frac{Q_n}{P_n}$ and $\frac{Q_{n-1}}{P_{n-1}}$.

If we notice that the distribution of types $G_\beta$ in $[\frac{Q_n}{P_n},\frac{Q_{n-1}}{P_{n-1}}]$ is identical to the distribution of types in $[\frac01,\frac11]$ we conclude that there is a certain relationship of hyperbolic self-similarity between sizes of stairsteps between $\frac{Q_n}{P_n}\;\&\;\frac{Q_{n-1}}{P_{n-1}}$ and sizes of stairsteps between $\frac01\;\&\;\frac11$: there is a subtle underlying law, $(F-B)$ based --hence hyperbolic in nature-- that rules the way in which the size of stairsteps decreases.

\section{Self-similarity of the underlying Set $\Omega$}

In the last section we saw that vertical sizes in the staircase decrease hyperbolically and horizontal sizes Euclideanly. We also saw that there was a particular aspect in which we could study the shrinking of intervals of resonance according to a hyperbolic law as well. Such ubiquitous hyperbolic changes point to a question that arises in a natural way: is the underlying set $\Omega$ hyperbolically self-similar?

We will consider the set $\Omega$ underlying the circle-map. We have two intervals of resonance $I^1_1$, $I^1_2$ associated with $[\frac01,\frac12]$ and $[\frac12,\frac11]$, respectively, {\bf via the staircase} (see Sec. 4.2). We have four intervals $I^2_i,\;i=1,...,4$ associated (via the staircase) respectively with segments $[\frac01,\frac13]$, $[\frac13,\frac12]$, $[\frac12,\frac23]$ and $[\frac23,\frac11]$ of the $(F-B)_2$ partition. In general we have $I^k_i,\;i=1,...,2^k$ associated with the $2^k$ segments in the $(F-B)_k$ partition. Segments in the $(F-B)_k$ partition decrease, when $k$ grows, in a hyperbolically self-similar way, and we want to compare sizes of segments $I^k_i$ with sizes of segments in the corresponding $(F-B)_k$ partition.

Figs. $2,3,4$ and $5$ show, for $k=3,4,5$ and $6$, sizes of $I^k_i$ in the horizontal axis plotted against the corresponding $(F-B)_k$ sizes in the vertical axis. The fact that each comparative figure shows a straight line indicates that $(F-B)_k$ segments and $I^k$ segments have sizes proportional to one another, the constant of proportionality $m_k$ being the slope of said straight line. It remains to show the law governing the change of these slopes $m_k$ when $k$ grows. A look at Fig. $6$ suggest that $m_k$ grows {\bf linearly} with $k$.

So $\Omega$ is hyperbolically self-similar. If we took {\bf another} circle-map, would the corresponding $\Omega$ be hyperbolically self-similar as well? Would the $\Omega$ underlying $q=g(-H)$ be hyperbolically self-similar? We conjecture that the answer to both questions is "yes": provided that the staircase fulfills equation (\ref{fases}) and has the $(F-B)$ arrangement in the vertical axis, we conjecture that the underlying $\Omega$ should be hyperbolically self-similar.

\section{Conclusions}
The leading subject in Number Theory  of approximating irrational numbers by rational ones can be tackled precisely when real numbers are expressed as continued fraction expansions. There is a partition of $I=[0,1]$ naturally associated with this expansion: the $(F-B)$ partition,... in much the same way as the decimal expansion of real numbers is naturally associated with the decimal partition of $I$ in ten segments of equal length, each of the latter in another ten... and so on. This $(F-B)$ partition is, in turn, naturally associated with the hyperbolic measure in $\Bbb H$.

Now, in order to tackle the problem of approximating an irrational $i=[a_1,...,a_n,...]$ by rationals $[a_1,...,a_n]=\frac{Q_n}{P_n}$, Jarn\'{\i}k classified irrationals in $J_\beta$ classes, $\beta\geq 2$, according to the corresponding degree of approximation --i.e. to the speed of convergence of $\frac{Q_n}{P_n}$ to $i$.

In order to study Cantor staircases in physics --forced pendulum, magnetization, etc.-- showing the $(F-B)$ arrangement for intervals $I(\frac{Q}{P})$, a natural connection with Number Theory appears, precisely due to the ubiquitous presence of the $(F-B)$ partition. But when closely examining the behaviour of these staircases, we were forced to considerably refine the $J_\beta$ nested classes into the $G_\beta$ disjoint ones.

{\bf We are saying that problems in empirical physics produced a refinement of key tools in Number Theory.}

The properties of these $G_\beta$, $\beta\geq 2$, allowed us to extract theoretical and practical information about the multifractal spectrum of such cantordusts $\Omega$ underlying Cantor staircases in physics, and about the nature of the self-similarity of said stircases.

\newpage

{\bf \Large References}

Bak, P. [1986] "The Devil's staircase," {\it Phys. Today}, December 1986, 38-45.

Bruinsma, R. and Bak, P. [1983] "Self-similarity and fractal dimension of the devil's staircase in the one-dimensional Ising model," {\it Phys. Rev.} {\bf B27}(9), 5924-5925.

Cvitanovic, P., Jensen, M., Kadanoff, L. and Procaccia, I. [1985] "Renormalization, unstable manifolds, and the fractal structure of mode locking," {\it Phys. Rev. Lett.} {\bf 55}(4), 343-346.

Falconer, K. [1990] {\it Fractal Geometry} (John Wiley and Sons, Chichester-New York).

Grynberg, S. and Piacquadio, M. [1995] "Hyperbolic geometry and multifractal spectra. Part. II," {\it Trabajos de Matem\'atica 252, Instituto Argentino de Matem\'atica, CONICET.} 

Halsey, T., Jensen, M., Kadanoff, L., Procaccia, I. and Shraiman, B. [1986] "Fractal measures and their singularities: The characterization of strange sets," {\it Nucl. Phys. B, Proc. Suppl. 2}, 513-516. 

Piacquadio, M. and Grynberg, S. [1998] "Cantor staircases in physics and diophantine approximations," {\it International Journal of Bifurcation and Chaos} {\bf 8}(6), 1095-1106.

Series, C. [1985] "The modular surface and continued fractions," {\it Journal of the London Mathematical Society} {\bf 31}(2), 69-80.


\clearpage

{\bf\Large Figure Captions}

\begin{figure}[h]
\begin{center}
\includegraphics[width=1.2\textwidth]{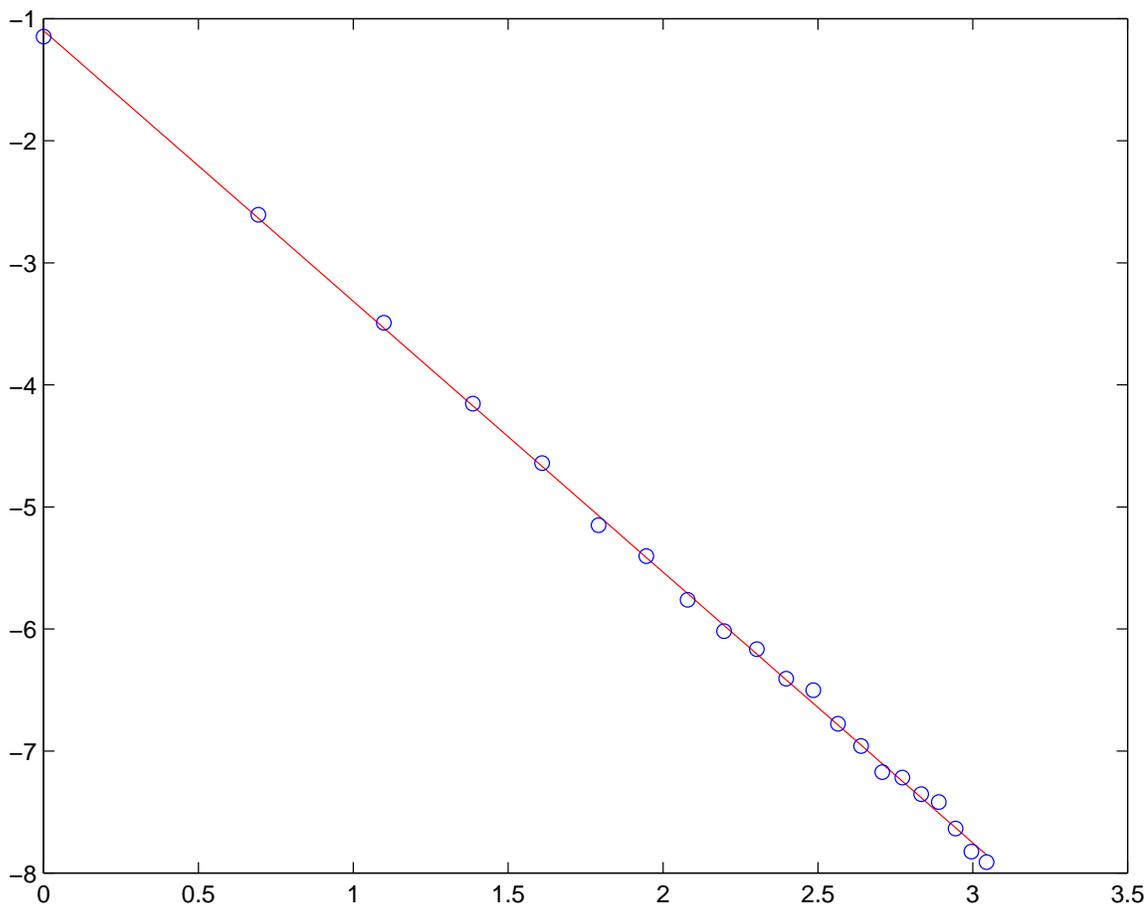}
\end{center}
\caption{Vertical variable $Y=\log(\Delta^*\omega(\frac{Q}{P}))$ plotted against horizontal variable $X=log(P)$ for the circle map staircase $W=g(\omega)$. Here $\Delta^*\omega(\frac{Q}{P})$ is the average taken on intervals $\Delta\omega(\frac{Q}{P})$ corresponding to the same value of $P$.}
\end{figure}

\clearpage

\begin{figure}[h]
\begin{center}
\includegraphics[width=1.2\textwidth]{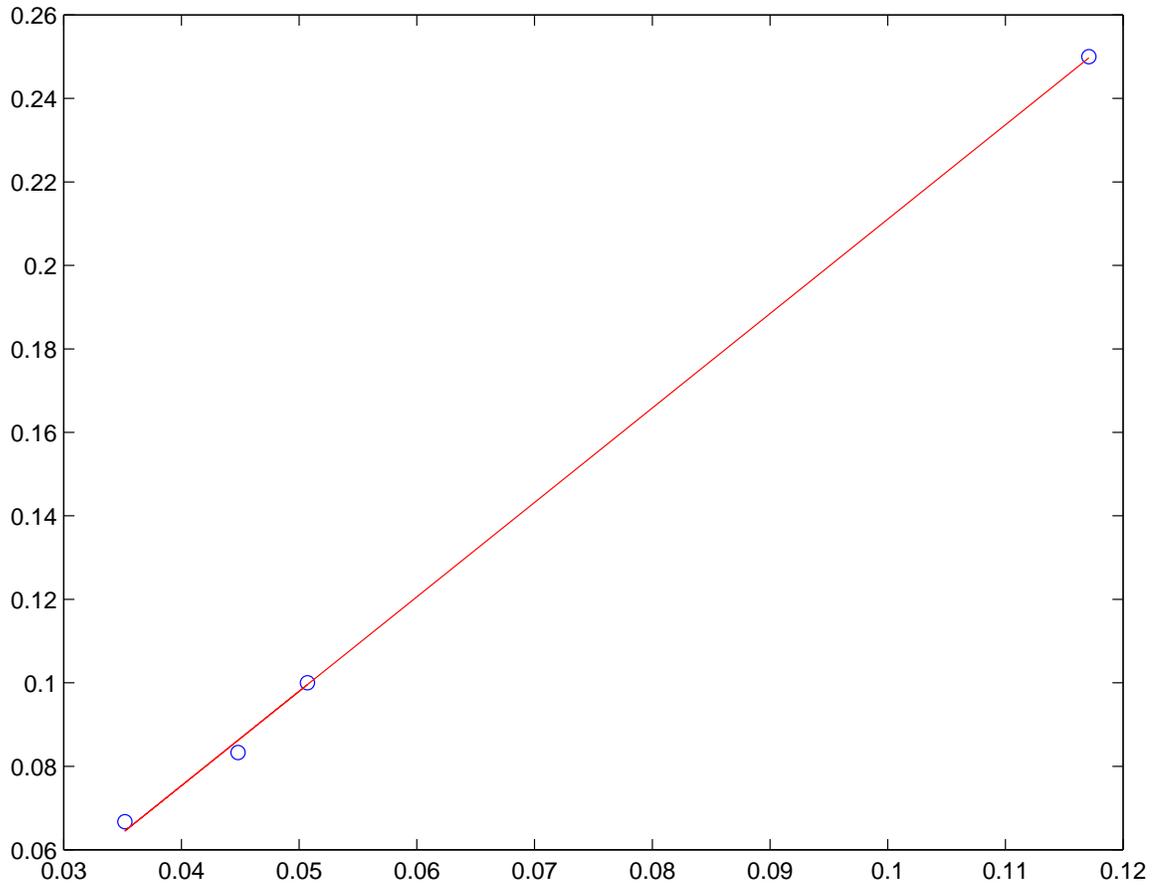}
\end{center}
\caption{Horizontal variable $X$ is the size of $I^3_i$ for $i=1,2,3,4$. Variable $Y$ is the size of the corresponding $(F-B)_3$ segment. The {\it other} four  $(i=5,6,7,8)$ out of a total of $2^3$ points coincide exactly with the four points shown in the figure.}
\end{figure}

\clearpage

\begin{figure}[h]
\begin{center}
\includegraphics[width=1.20\textwidth]{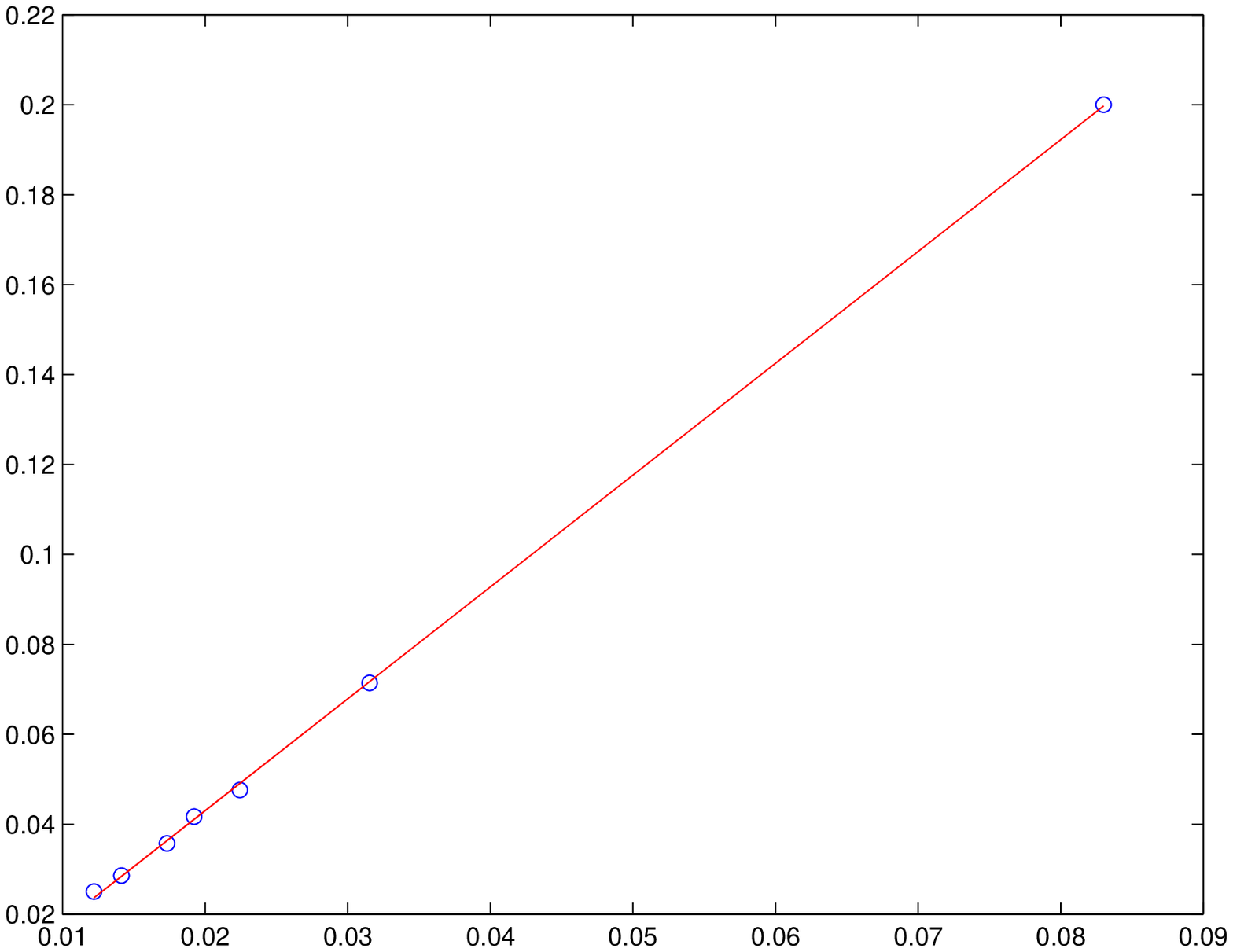}
\end{center}
\caption{Variable $X$ is the size of $I^4_i$; variable $Y$ is the size of the corresponding $(F-B)_4$ segment.}
\end{figure}

\clearpage

\begin{figure}[h]
\begin{center}
\includegraphics[width=1.20\textwidth]{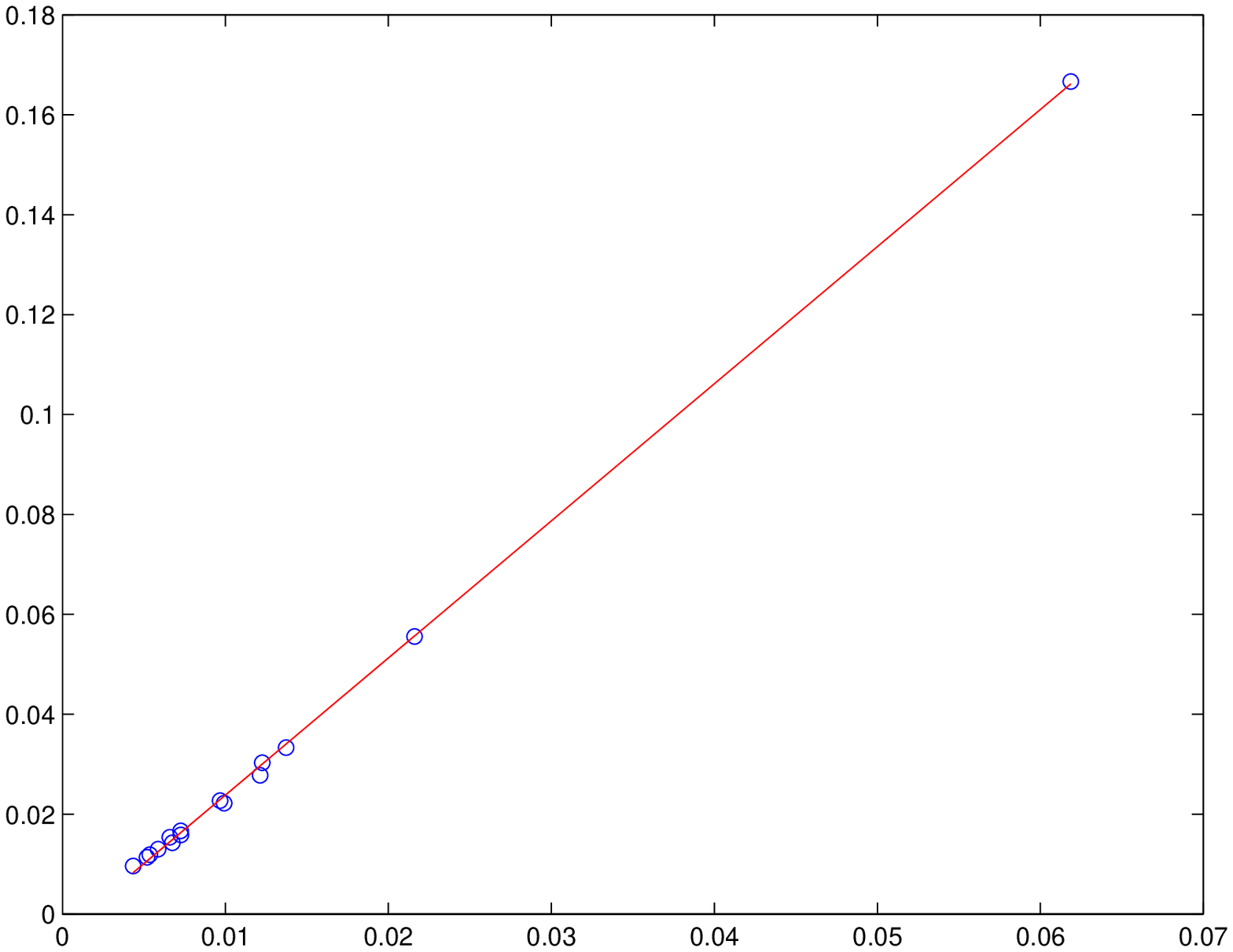}
\end{center}
\caption{Variable $X$ is the size of $I^5_i$; variable $Y$ is the size of the corresponding $(F-B)_5$ segment.}
\end{figure}

\clearpage

\begin{figure}[h]
\begin{center}
\includegraphics[width=1.20\textwidth]{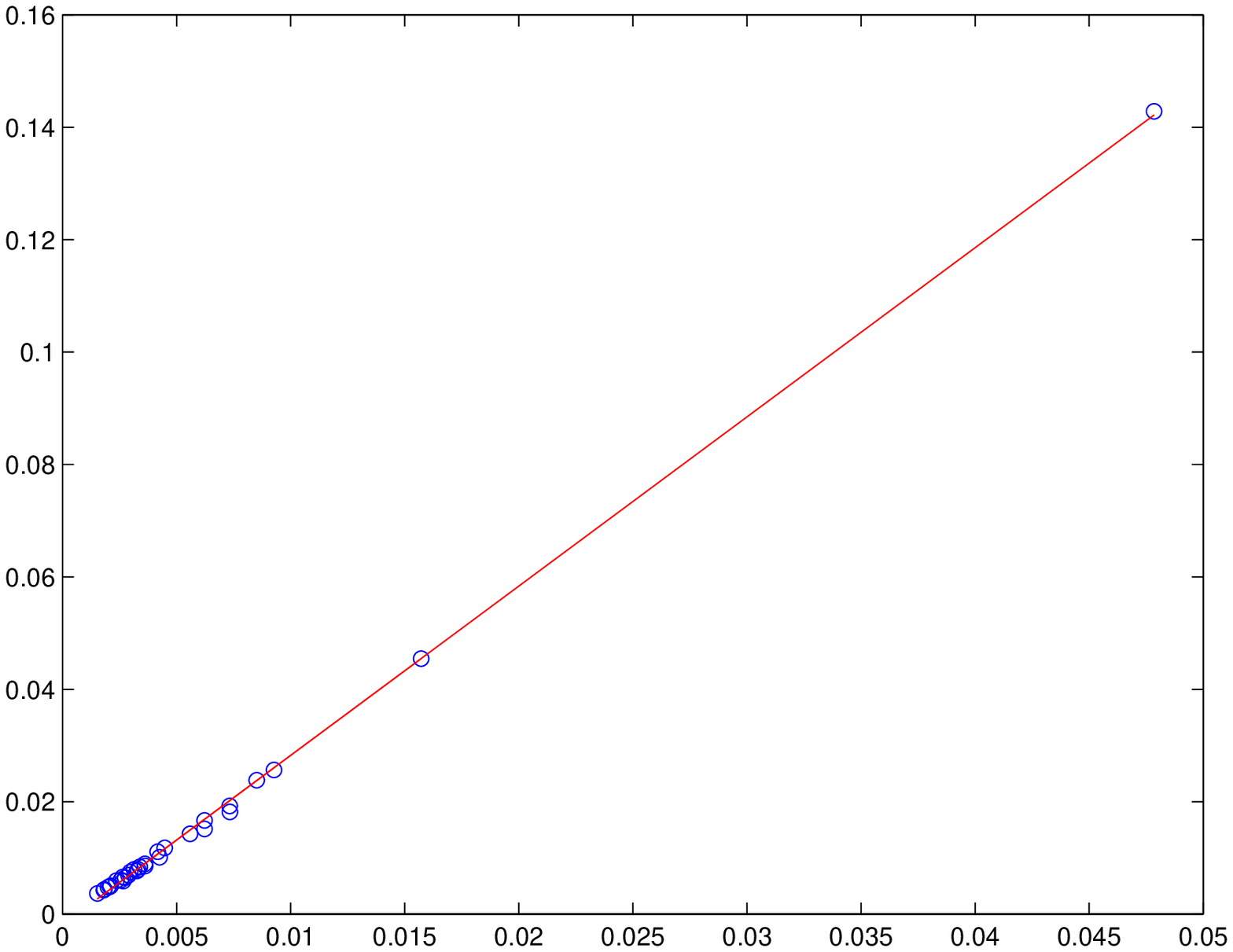}
\end{center}
\caption{Variable $X$ is the size of $I^6_i$; variable $Y$ is the size of the corresponding $(F-B)_6$ segment.}
\end{figure}

\clearpage

\begin{figure}[h]
\begin{center}
\includegraphics[width=1.20\textwidth]{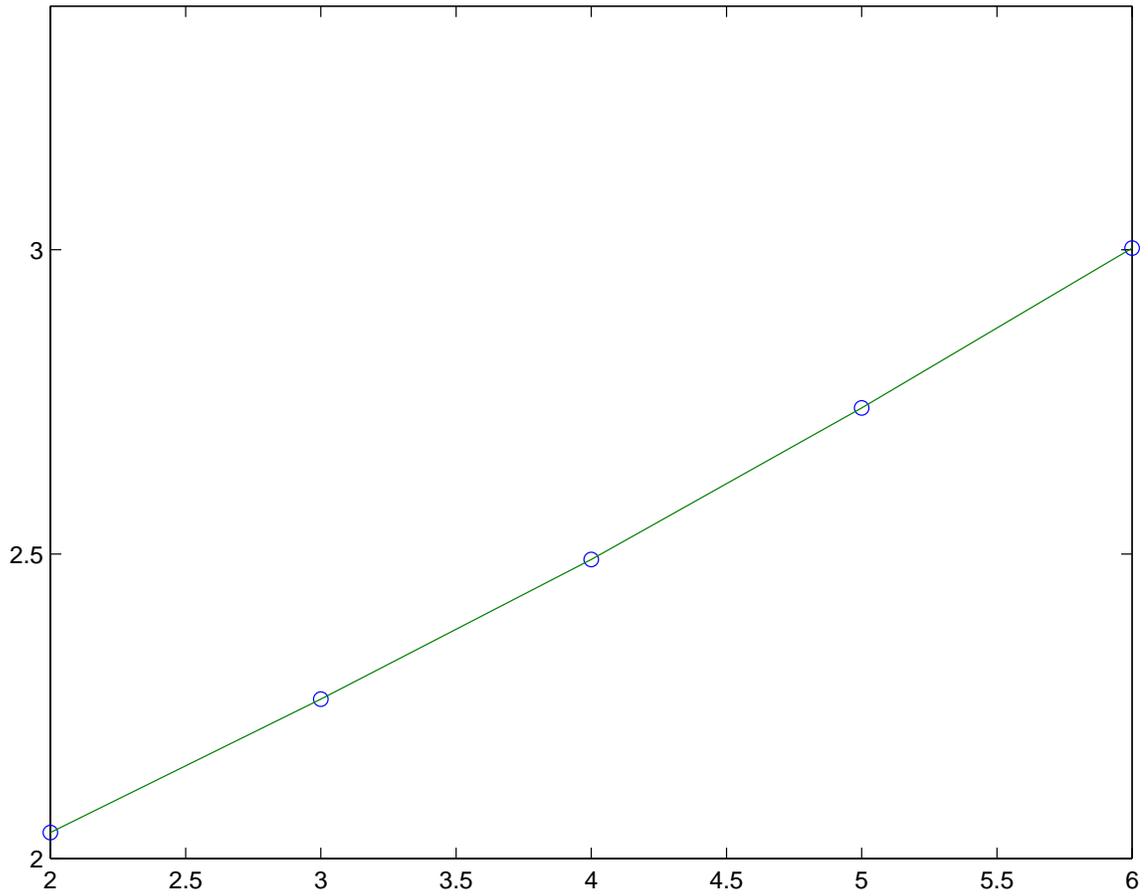}
\end{center}
\caption{Slopes $m_k$, $k=3,4,5$ and $6$ of lines in figures $2,3,4$ and $5$ plotted  against $k$. We have added $m_2$, the slope of the (trivial) line joining the two points for the second approximation of $\Omega$ and the corresponding $(F-B)_2$ segments.}
\end{figure}

\end{document}